# Crystal Structure and Chemistry of Topological Insulators

R.J. Cava[1], Huiwen Ji[1], M.K. Fuccillo[1], Q.D. Gibson[1] and Y.S. Hor[2]

[1]Department of Chemistry, Princeton University, Princeton NJ 08544

[2]Department of Physics, Missouri University of Science and Technology, Rolla MO 65401

**Abstract**

Topological surface states, a new kind of electronic state of matter, have recently been observed on the cleaved surfaces of crystals of a handful of small band gap semiconductors. The underlying chemical factors that enable these states are crystal symmetry, the presence of strong spin orbit coupling, and an inversion of the energies of the bulk electronic states that normally contribute to the valence and conduction bands. The goals of this review are to briefly introduce the physics of topological insulators to a chemical audience and to describe the chemistry, defect chemistry, and crystal structures of the compounds in this emergent field.

**Introduction**

Physicists have recently been pleasantly surprised by the prediction and then observation of new electronic states with intriguing properties, called "topological surface states".[1-5] The electrons in these states, which are found on the surfaces or the edges of small band gap semiconductors where the bulk of the material is electronically insulating (with caveats, as described below), display metallic conductivity with unusual characteristics. The inclusion of heavy atoms in the host bulk materials, causing spin orbit coupling to be high, is a critical factor in making the electronic states on the surfaces chiral, which imparts many of their unique properties. The chirality makes the surface state electrons immune from direct backscattering when encountering surface or edge defects. (Because the surface electron spins are locked perpendicular to their momentum, the spin orientation would have to flip in order for the electrons to change direction by 180 degrees, which is forbidden.) Their energy quantization is more Dirac-like (i.e. photon-like) than bulk-electron-like. These states have inspired predictions of new kinds of electronic devices and exotic physics, including proposals for detecting a long sought neutral particle obeying Fermi statistics called the "Majorana Fermion".[6, 7] Topological surface states were first observed by transport measurements on the edges of thin buried layers of HgTe in quantum wells fabricated by molecular beam epitaxy (MBE),[4] and by photoelectron spectroscopy on the surfaces of bulk crystals of Sb-doped Bi, $Bi_{0.9}Sb_{0.1}$.[5] In principle, the discovery of a new type of electronic state is what is critically important, regardless of whether experimental observations are made on MBE-fabricated quantum wells, or on the surfaces of bulk crystals. The fact that the exotic electronic states can be observed on the cleaved surfaces of easily synthesized bulk crystals, however, means that a large number of experimentalists have been able to contribute to progress in understanding them, and has led to the explosion of interest in topological insulators in recent years, making the bulk-crystal-based materials of particular importance. In fact one can view the bulk-based topological insulators as "ideal materials" for condensed matter physics research: the physics is new and exciting, with plenty of opportunity for both new theory and experimental discovery, and the chemistry is "simple". In addition, the crystal structures are not too complex, and large, good quality crystals can be grown in many laboratories world-wide.

There have already been many physics-based reviews of topological insulators in both the technical and popular scientific literature,[8-12] and the physics of topological insulators has been addressed from the chemical perspective.[13] Although the materials themselves have been critical to the development of the field from the very beginning, there is currently no review treating their basic structure and chemistry. We address those aspects of the materials in the current review for the cases where confirmed experimental observation of the surface states has been made. After the prediction of the existence of topological surface states on $Bi_{1-x}Sb_x$ alloys that started the research in the bulk crystal TIs,[3] many predictions of their existence on the surfaces of different kinds of crystals have been made. These predictions have been widely possible due to the fact that the current generation of electronic structure calculation programs can describe the bulk and surface electronic states of main-group-based materials. This has inspired electronic-structure-calculation-based predictions of new TIs that have ranged from calculating and screening the partial electronic structures of 60,000 compounds found in the ICSD database without human intervention, a study that at this time has not yet been formally published,[14] to predicting the detailed behavior for individual compounds, published in high impact journals.

From the perspective of solid state chemistry, the first order requirements for bulk topological insulators are relatively simple. First, the electron count of the compound must yield a semiconductor, either by simple electron counting or the Zintl concept. Next, the electronegativies of the elements whose orbitals are involved near the Fermi energy must be close to each other, resulting in strong covalency. The bonding should result in some mixing of the states that would normally be present at the bottom of the conduction band (i.e., states from the more electropositive element) into the top of the valence band, and states from the less electropositive element into the bottom of the conduction band. This would normally lead to a semimetal (a metallic conductor due to the presence of a continuous manifold of energy states that arises when the conduction band minimum dips below the valence band maximum for electrons at particular wave vectors), but the presence of spin orbit coupling splits the energies of the bands, analogous to the way crystal fields split orbital energies, and a small band gap semiconductor results instead. If the band gap is too big, then there is no mixing of the states, and if the overlap of states is too strong then the spin orbit coupling will not be enough to open a band gap. So a balance is needed. More detailed physics such as the parity and degeneracy of the bands is also important.

The review is organized as follows. First, HgTe and elemental Bi and Sb are described. Then the larger and so far most frequently studied class of compounds based on the Tetradymites $Bi_2Te_3$ and $Bi_2Se_3$ is described; the next structurally simplest family, based on $TlBiSe_2$, follows. After that, we describe two different more complex families of homologous series, the first the infinitely adaptive series $(Bi_2)_n(Bi_2Se_3)_m$, which has given rise so far to two "topological insulators" $Bi_4Se_3$ and BiTe. (They may actually be semimetals. Such compounds may still display topological surface states even though they will not be insulating in the bulk.) The description of a series based on $GeTe-Bi_2Te_3$ or $PbTe-Bi_2Te_3$ compounds, typified by $GeBi_2Te_4$ and $GeBi_4Te_7$, follows. These materials display common structural and chemical characteristics.

Finally, (Pb,Sn)Se and SnTe, which are examples of the relatively new class of topological materials that have been given the chemically confusing name "topological crystalline insulators"[15] (they are all crystalline after all) are described. Since progress in the field is rapid, other classes of materials displaying topological surface states may appear before this review is published. What we describe here is current as of the beginning of 2013. A brief description of the defect chemistry of the compounds and general comments conclude the review.

**Structures and Chemistry**

HgTe

Mercury telluride is one end member of the much studied $Hg_{1-x}Cd_xTe$ series of small band gap semiconductors.[16] This solid solution has been of interest for decades for infrared detectors. This is the canonical example of a solid solution where the decrease of the average electronegativity of the main group "metallic" element in the $Hg_{1-x}Cd_xTe$ series results in a change from semimetal behavior for HgTe to a 1.5 eV semiconductor for CdTe. (All the compounds described are at most polar covalent compounds, not ionic compounds, though we will use formal charges to describe the ions for convenience. For example, for the II-VI compounds described in this section, the Pauling electronegativities are 1.7, 2.0 and 2.1 for Cd, Hg and Te, respectively; so the compounds are not ionic.) The $Hg_{1-x}Cd_xTe$ materials are direct band gap semiconductors with the minimum energy difference between valence and conduction bands at the Γ point ((0,0,0) in reciprocal space) in the Brillouin zone. (The Brillouin zone is the volume of reciprocal space that encompasses all the unique wavevectors for electrons in the solid. At the Γ point, the electron waves are very long compared to crystallographic unit cell dimensions.) HgTe is therefore normally a metallic conductor. When a thin layer of HgTe is sandwiched between layers of CdTe, then the electrons are confined to a two-dimensional conductive sheet. This causes the electrons to travel in spin polarized chiral states at the edges of the buried layer in the fabricated quantum wells. The experiments on HgTe showed that a critical minimum layer thickness is needed (on the order of 60 Å) for the observation of the edge states.[4]

HgTe has the Zincblende structure (Fig. 1). This structure is common for group IV semiconductors such as Si and Ge, III-V compound semiconductors such as InP and GaAs, and II-VI semiconductors such as ZnS (Zincblende) and CdSe. The crystal structure has space group $F\bar{4}3m$ (#216), and is a derivative of diamond. There are two symmetry independent atomic sites per cell, both occupied by the same element in C, Si and Ge, but by different elements in the compound semiconductors. In HgTe, the cation Hg occupies the Wyckoff *4a* (000) positions (with equivalents at the FCC translations 0,½,½; ½,0,½; and ½,½,0) in the cubic cell, and Te, the anion, is found in one half of the potential tetrahedral interstices, Wyckoff positions *4c* (¼,¼,¼) (plus the FCC equivalents). With a unit cell parameter at room temperature of 6.460 Å, the Hg-Te bond length in the Hg-Te tetrahedra is 2.797 Å.[17]

Elemental Bi and Sb, and $Bi_{1-x}Sb_x$

Elemental Bi and Sb are isostructural (Fig. 2a), with interesting crystal structures.[18, 19] Their ambient pressure structures are rhombohedral, space group $R\bar{3}m$ (#166). There is one symmetrically independent atomic site per cell, in Wyckoff position *6c*, coordinates (0,0,0.23389) for Bi and (0,0,0.23349) for Sb, plus atoms equivalent by inversion and the rhombohedral centering translations (2/3,1/3,1/3) and (1/3,2/3,2/3). (The positions of equivalents will not be given after this point. All of the rhombohedral phases will be described in the hexagonal coordinate system. All the structural information presented is for room temperature.) The hexagonal cell parameters are $a$ = 4.546 Å and 4.3084 Å, and $c$ = 11.862 Å and 11.274 Å for Bi and Sb, respectively. These are six layer unit cells, with the close packed layers of Bi or Sb in the sequence AB CA BC. The unit cell is actually made of three Bi or Sb bi-layers (emphasized by the use of spaces in the stacking sequence presented in the previous sentence). The c/a ratios are 2.61 and 2.62 respectively, much less than the c/a ratio for 6 layers of an ABCABC-stacked FCC material such as Cu, where the ratio is 4.9. This is because the Bi(Sb) bi-layers are almost collapsed into being a single layer, making them smaller perpendicular to the planes and larger within the planes than the ideal case. The bonds are stronger within the bi-layers than between adjacent bi-layers. Within the close-packed plane of the individual Bi (Sb)-layers, the interatomic distances are 4.546 Å (Bi) and 4.308 Å (Sb). Between layers in the bi-layer, the three Bi-Bi and Sb-Sb bonds are quite short, being 3.071 Å and 2.908 Å respectively. Between adjacent bi-layers the bonds are substantially longer, 3.529 Å (Bi) and 3.355 Å (Sb). The layered crystal structure of these elements is a consequence of the electronic configurations of elemental Bi and Sb ($6p^36s^2$ and $5p^35s^2$ respectively).[20] Each Bi or Sb atom is bonded to 3 nearest neighbors to close its valence shell, and the lone pair 6s or 5s electrons squeeze the 3 Bi-Bi (Sb-Sb) bonds stereochemically to one side of the central atom.

The $Bi_{1-x}Sb_x$ solid solution is single phase from x = 0 to 1 with no long range or short range order of the elements yet detected,[21, 22] implying that there is no preference for local self or heteroatom bonding for Bi and Sb. The solid solution is interesting electronically because different electronic bands change their energies as a function of composition. Bi and Sb are both semimetals. The origins of the overlapping bands yielding the semimetal behavior are different for elemental Bi and elemental Sb. The bands change energy dramatically with composition, and at x values between 0.07 and 0.23, a true semiconductor is observed.[23] In this narrow semiconducting composition regime the states normally giving rise to the valence and conduction bands are inverted in energy, a requisite for the formation of the surface states.

The surface states are, luckily, found on the good cleavage surface (Fig. 2b), variously called the "basal plane", the "111 plane", or the "001 plane". This plane has three-fold rotational symmetry and the topological surface states cross the Fermi energy the requisite odd number of times. The Fermi surface for the electrons in surface states is rather complicated, however, making the $Bi_{0.9}Sb_{0.1}$ alloy less than an ideal material for studying topological surface states at this time.

Tetradymites

Due to the importance of this group of compounds in the field, quite a few variants of Tetradymite have been studied from the TI perspective. The compounds are based on the stacking of three "quintuple layer" building blocks to yield a crystallographic cell with rhombohedral symmetry (Fig. 3a). For $Bi_2Se_3$, for example, the layers are –[Se(2)-Bi-Se(1)-Bi-Se(2)]$_0$–[Se(2)-Bi-Se(1)-Bi-Se(2)]$_{1/3}$–[Se(2)-Bi-Se(1)-Bi-Se(2)]$_{2/3}$–. (The subscripts indicate the fractional translation of the quintuple layer sandwich along $z$ in the hexagonal unit cell). Critical to their usefulness in the field has been their excellent (001) plane cleavage (Fig. 3b), due to the presence of the relatively weak van der Waals bonds between adjacent quintuple layer building blocks. This cleavage exposes the outer Te(2) or Se(2) planes of the building blocks, which then host the topological surface states (Fig. 3(b)).

The compounds crystallize in space group R$\bar{3}$m (#166). For $Bi_2Se_3$,[24] the crystal structure parameterization is $a$ = 4.143 Å, $c$ = 28.636 Å; Bi position 6c (0,0,0.3985); Se(1) position *3a* (0,0,0); Se(2) position 6c (0,0,0.2115). For $Bi_2Te_3$,[25] it is $a$ = 4.395 Å, $c$ = 30.440 Å; Bi *6c* (0,0,0.4005); Te(1) 3a (0,0,0); Te(2) 6c (0,0,0.2097). And, finally, for $Sb_2Te_3$,[26] it is $a$ = 4.264 Å, $c$ = 30.458 Å; Sb *6c* (0,0,0.3988); Te(1) *3a* (0,0,0); Te(2) *6c* (0,0,0.2128). $Sb_2Se_3$ (unfortunately, see below) does not form with the same structure type. None of the simple binary Tetradymite phases has yet to be made insulating through small amounts of doping: $Bi_2Te_3$ made at the stoichiometric ratio is naturally p-type. Although it can be manipulated to display n- or p- type behavior, it is difficult to control the stoichiometry well enough to make it display low carrier concentrations. $Bi_2Se_3$ is naturally n-type, though it can be driven p-type by small amounts of alkaline earth doping. $Sb_2Te_3$ is strongly p-type, with high carrier concentrations, and has not yet been made n-type by doping.[27, 28] Please refer to the defect chemistry section of this review for further details.

To try to induce bulk insulating behavior in the Tetradymites, more complex ternary and quaternary phases have been studied. The fact that $Sb_2Se_3$ forms in a different structure type means that the full ternary $Bi_{2-y}Sb_ySe_3$ and full quaternary $Bi_{2-y}Sb_yTe_{3-x}Se_x$ solid solutions do not form for all compositions from $0 \leq y \leq 2$ and $0 \leq x \leq 3$. The quaternary solid solution has long been studied in the optimization of room temperature thermoelectric materials.[27, 28] In the ternary $Bi_{2-y}Sb_ySe_3$ solid solution, the y limit is less than 0.5.[29] There appears to be no report of long or short range Sb/Bi ordering in this series. $Bi_2Se_3$ and $Bi_2Te_3$ are isostructural and the full single phase solid solution series $Bi_2Te_{3-x}Se_x$ is known for $0 \leq x \leq 3$.[24] In this case there is a critical (from the TI perspective) long range ordering scheme within the solid solution (Fig. 4). Because the Te(1)/Se(1) site in the middle of the quintuple layer is more ionic in character (i.e., it is bonded to Bi in both the plane above and the plane below), that site fills first as the more electronegative Se is substituted for Te in $Bi_2Te_3$, filling virtually perfectly until y = 1.[24, 30, 31] In other words the Tetradymite at the formula $Bi_2Te_2Se$ (BTS) is a crystallographically ordered compound and may be considered a distinct phase; thermodynamic anomalies at this composition have been observed, for example.[32] The structure of $Bi_2Te_2Se$ has been parameterized in space group $R\bar{3}m$ (#166) as is $a$ = 4.3067 Å, $c$ = 30.0509 Å; Bi position *6c* (0,0,0.3968); Se position *3a* (0,0,0); Te position *6c* (0,0,0.2116).[31] A recent structure determination has found about 4% mixing of Se on the Te sites (an electrically neutral defect) in Bridgman-grown crystals, by both STM and conventional XRD analysis.[31] The stacking sequence for $Bi_2Te_2Se$ is then: –[Te-Bi-Se-Bi-Te]$_0$–[Te-Bi-Se-Bi-Te]$_{1/3}$–[Te-Bi-Se-Bi-Te]$_{2/3}$–. This compound has proven to be quite important in the TI field thus far because it can be made as a bulk insulator (i.e. with a resistivity of 1-10 ohm-cm at 4 K, compared to the $10^{-3}$ ohm-cm values typically found for the binary Tetradymites $Bi_2Se_3$ and $Bi_2Te_3$), with nearly balanced donor and acceptor states, facilitating the observation of the transport of the topological surface electrons. For x > 1 in $Bi_2Te_{3-x}Se_x$, the selenium also partially occupies the Te sites on the outer layer of the quintuple layer sandwich, and a disordered solid solution phase forms with significant Te/Se mixing on the cleaved basal plane surface that supports the surface states.[24] Materials with x = 2, for example, can be written as $Bi_2(Te_{0.5}Se_{0.5})_2Se$. These Se-rich phases are naturally n-type and can be made p-type by substitution of Sb for half of the Bi, with the n to p crossover near $(Bi_{0.5}Sb_{0.5})_2(Te_{0.5}Se_{0.5})_2Se$.[33] Topological surface states have also been observed on the basal plane surfaces of these complex disordered quaternary $(Bi_{1-x}Sb_x)_2(Te_{1-x}Se_x)_3$ solid solutions,[34] implying that structural disorder without charge disorder (the mixed ions have the same formal charge) is not necessarily a game-ending problem for the stability of topological surface states.

As research on topological insulators has become more advanced, experimentalists hope to observe the character of the surface states in more detail. Of particular interest is the behavior of electrons at the "Dirac point", the energy at which the Fermi surface of the surface electrons decreases in radius to become a point rather than a circle. Due to the delicate balance between donors and acceptors needed to hit this energy exactly, it is difficult to achieve for bulk crystals.

When the hole or electron counts in the bulk crystal are far off, say in the $10^{18}$ cm$^{-3}$ range, then the Fermi energy is in the bulk valence band or bulk conduction band, obscuring the surface state contribution. When the bulk carrier concentration is smaller, in the $10^{17}$ cm$^{-3}$ range or less, then the Fermi energy can fall in the surface state bands, however these low carrier concentrations are harder to obtain. Using the method of "ionic liquid gating", most successfully employed in materials physics to control the carrier concentration in thin superconductor layers,[35] researchers have been able to move the Fermi energy of the surface electrons closer to the Dirac point by design, through draining electrons from or injecting electrons into the surface layer of crystals through the application of an electric field. This is where ternary $Bi_2Te_2Se$ and quaternary $(Bi,Sb)_2(Te,Se)_3$ Tetradymites have been most important in TI research so far, as their bulk carrier concentrations can be made low enough to put the Fermi energy within the surface state bands and they have been amenable to liquid gating experiments that have gotten the Fermi energy quite close to the Dirac point.[36, 37]

There is, however, a complication for Tetradymites. For $Bi_2Se_3$, where the bulk electron concentration is unfortunately still too high, the Dirac point for the surface states falls at an energy that is distinctly well separated from the energies of the top of the bulk valence band and the bottom of the bulk conduction band. Electrons at the Dirac point in the surface states are therefore not strongly interfered with by electrons in the bulk states. For $Bi_2Te_3$ the Dirac point is clearly at an energy lower than the top of the bulk valence band, and thus even if a low hole or electron concentration could be obtained to place the Fermi energy near the Dirac point, the bulk valence band states would interfere strongly with the detection and characterization of the surface state electrons. For $Bi_2Te_2Se$, although the bulk carrier concentration can be decreased to the $10^{16}$ cm$^{-3}$ level, the Dirac point energy appears to be near the energy of the top of the valence band, suggesting that the bulk states will interfere. It is not yet known in detail how correct these relative energies are, and thus it is unclear whether this characteristic will limit the usefulness of this material for the ultimate transport experiments. Although experimental physicists can distinguish surface from bulk electron conductivity, it may not prove possible to inject enough electrons into the surface layer of the crystals to get to the surface state Dirac point. For quaternary $(Bi,Sb)_2(Te,Se)_3$ Tetradymites the Dirac point appears to be further from the top of the valence band for Bi to Sb ratios near 1:1,[34] but this has not yet led to the observation of any special surface state transport properties.

The problem of the Dirac point energy position has led to a search for additional Tetradymite compounds where the energy of the top of the bulk valence band is significantly lower than the energy of the surface state Dirac point. Another material has been found in this context, $(Bi,Sb)_2(Te,S)_3$. The chemical logic for looking at a Te-S-based compound rather than a Te-Se-based compound is that the higher electronegativity of S should decrease the absolute energy of the valence band, and thus bring it lower than the energy of the Dirac point, a reasoning that was shown to be correct. The crystal structure of the mineral "$Bi_2Te_2S$", the basis for this variant of Tetradymite, was considered by Linus Pauling in the 1970s.[38] He realized that the structure could not be stable at that composition due to the very strong strain imposed on the internal sulfur layer in the stacking sequence $-[Te-Bi-S-Bi-Te]_0-[Te-Bi-S-Bi-Te]_{1/3}-[Te-Bi-S-Bi-Te]_{2/3}-$, as, in analogy to the $Bi_2Te_2Se$ solid solution, the more electronegative anion (S) strongly prefers the middle layer in the quintuple layer Tetradymite building block. The strain arises from the very different ionic radii of S and Te, which would result in a large size mismatch of the hexagonal close packed S and Te layers. Pauling proposed that the stable compound formula for this Tetradymite should include S in the Te layer to relieve the strain (Pauling's proposal has S in the Te layer, and Te in the middle S layer as well, so a more complex formula of $Bi_{14}Te_{13}S_8$). A Tetradymite near this composition has been the subject of recent investigation as the host for topological surface states and has been shown to display a Dirac point that is far in energy above the top of the bulk valence band and also well separated from the bottom of the bulk conduction band.[39] Normally n-type, this compound also displays low carrier concentrations near the n to p crossover when Sb is substituted for about half of the Bi.

The crystal structure has been solved for $Bi_2(Te_{0.8}S_{0.2})_2S$,[39] and an interesting observation has been made about the effect of size mismatch on the crystal structure of Tetradymites (Fig. 5). As implied by the formula, the outer layers of the quintuple layer sandwich are occupied by an 80/20 ratio of Te to S. The crystal structure parameterization for $Bi_2(Te_{0.8}S_{0.2})_2S$ is $a$ = 4.196 Å, $c$ = 29.44 Å; Bi *6c* (0,0,0.39316); S(1) *3a* (0,0,0); Te(2) *6c* (0,0,0.21237), occupancy 0.8, S(2) *6c* (0,0,0.2266) occupancy 0.2.[39] The structure refinement clearly showed that the S atoms in the outer Te/S layer are closer to the Bi than the Te atoms are, consistent with a chemically reasonable Bi-S bond length. In other words, the layer of surface atoms that supports the topological surface states is randomly corrugated at the atomic scale as one passes over S and Te atoms at different heights. As is the case for the chemical disorder on this surface and the underlying layers present in the $(Bi,Sb)_2(Te,Se)_3$ ternary and quaternary phases, the positional disorder at the atomic level does not appear to kill the surface states in $Bi_2(Te_{0.8}S_{0.2})_2S$, as they have been observed in this compound. Finally, consistent with Pauling's deduction about the importance of strain in Tetradymites, the amount of S needed to stabilize the compound is smaller when smaller Sb is partially substituted for Bi. The formula of the chemically stable, high resistivity material with the well isolated Dirac point for this quaternary Tetradymite correspondingly has a stoichiometric Te to S ratio, with the formula $Bi_{1.1}Sb_{0.9}Te_2S$. Single crystals of this resistive material have been grown, and the Dirac point has been found by spectroscopy to be more than 100 meV above the top of the valence band, a favorable combination of properties. Thus far, however, just as for the $(Bi,Sb)_2(Te,Se)_3$ Tetradymites, there have been no reports about whether this compound is superior or inferior to the materials already known for use in ionic liquid gating studies.

There has also been recent interest in the $(Bi_{1-x}In_x)_2Se_3$ Tetradymite-type solid solution. The intermediate compositions between 0.3 eV band gap $Bi_2Se_3$ and 1.3 eV band gap $In_2Se_3$ [40] display an increasing band gap with increasing In content in a fashion analogous to what is found for $Hg_{1-x}Cd_xTe$. At some band gap size in the solid solution the $Bi_2Se_3$ topological surface states disappear. This is not surprising because as the band gap increases in size, the mixing of metal states in the valence band and selenium states in the conduction band decreases dramatically; this being the primary chemistry-based condition for the occurrence of topological insulators. Nonetheless the fact that the topological surface states disappear at some composition ($x \sim 0.1$) in $(Bi_{1-x}In_x)_2Se_3$ is of interest to physicists because this is one of the possible systems where such disappearance can be studied.[41]

TlBiSe$_2$, TlBiTe$_2$ and the TlBiSe$_{2-x}$S$_x$ series

Monovalent Tl can be an interesting ion in solids. It is sometimes found in ternary layered or cage compounds where intermediate size alkali (e.g. K) variants are also known, e.g. $KFe_{2-x}As_2$-$TlFe_{2-x}Se_2$ and $K_{0.3}WO_3$-$Tl_{0.3}WO_3$.[42, 43] On the other hand, Tl frequently is found in more unique Tl-based compounds that depend on Tl's large size, its 1+ charge, intermediate electronegativity, and lone pair configuration (its valence shell is $6p^1 6s^2$) for their stability. In the topological insulator family, Tl is so far known in the compounds TlBiSe$_2$ and TlBiTe$_2$, and in the solid solution TlBiSe$_{1-x}$S$_x$, which is one of the cases where a crossover in behavior between compounds that display topological surface states (x = 0) and others that do not occurs within the same structural and chemical family.[44-46] The Pauling electronegativities for the atoms involved in this series are 1.8, 2.0, 2.1, 2.5 and 2.6 for Tl, Bi, Te, Se and S, respectively. Tl is a relatively electropositive metal and S is relatively electronegative.

These compounds also crystallize in the rhombohedral space group $R\bar{3}m$ (#166), with the metal atoms in special positions: Tl 3a (0,0,0); Bi 3b (0,0,0.5) and the anions in positions with one variable coordinate, z: Se or Te 6c (0,0,0.25), S 6c (0,0,0.262).[47-49] The cell parameters are $a$ = 4.1041 Å (S), 4.24 Å (Se), and 4.527 Å (Te), and $c$ = 21.872 Å (S), 22.33 Å (Se), and 23.118 Å (Te). The ABCABC stacking of the close-packed layers (Fig. 6) is in the order -(S,Se,Te)-Bi-(S,Se,Te)-Tl-. Continuous bonding is found perpendicular to the layers; there are no van der Waals bonds present. Curiously, with all atoms in special positions, or nearly special positions (The S position in $TlBiS_2$ appears to have been determined to better precision than the Se and Te positions in $TlBiSe_2$ and $TlBiTe_2$; it is slightly off the midpoint between the metal layers. This could be specific to S or it could be that the same is true for Se and Te. Higher precision crystal structure determinations should be performed on those compounds.), there are no asymmetries in the reported crystal structures that reflect of the presence of the lone pairs on the $Bi^{3+}$ or $Tl^{1+}$. In addition, it is not clear at this point how well the Tl/Bi ordering has actually been determined, since the structures have been solved by conventional X-ray diffraction and $Tl^{+1}$ and $Bi^{+3}$ are nominally identical in electron count. (Each has 80.) They are therefore indistinguishable by X-rays. The fact that there is no asymmetry in the crystal structure, if that is correct, suggests that there could be significant Tl/Bi disorder in the metal sites. In spite of the absence of van der Waals bonding, the crystals form very nice (001) plane cleavage surfaces, which is where the topological surface states have been observed. There has of yet been no report of the characterization of the cleavage surfaces by scanning tunneling microscopy (STM) or photoemission electron microscopy (PEEM) measurements so the identity of the terminating atomic layer (or layers) on the cleaved surfaces is not currently known.

$Bi_4Se_3$, BiTe and the $(Bi_2)_n(Bi_2X_3)_m$ (X=Se and Te) infinitely adaptive series

This is a large and crystallographically interesting series of compounds found between elemental Bi metal and the ionic-like end members $Bi_2Te_3$ and $Bi_2Se_3$ (and between Sb and $Sb_2Te_3$). It has been well studied from the crystallographic perspective and is an excellent example of the concept of the infinitely adaptive series in solid state chemistry.[50, 51] The idea is that very stable structural components, in this case $Bi_2$ bilayers and $Bi_2Se_3$ or $Bi_2Te_3$ quintuple layers, may in rare cases maintain their identities and stack in arbitrary integer ratios, forming an infinite series of distinct compounds between the pure end members. In the TI-relevant systems, an "infinitely adaptive series" of compounds with the formulas $(Bi_2)_n(Bi_2Se_3)_m$ (for the selenides) for arbitrary $n$ and $m$ (Fig. 7) is formed. Distinct compounds are formed due to the stability of the individual building blocks, and, critically, because bonding between unlike blocks is preferred over bonding between identical ones. The compounds display fully ordered, not random, stacking arrangements.

Typical of such series, although the end members (elemental Bi and the compounds $Bi_2Se_3$ or $Bi_2Te_3$ for example) are relatively easy to grow as single crystals, the more complex members of the series are quite challenging to isolate and grow as clean crystals due to the limited temperature range over which they are stable compared to the simpler members of the series. For the materials of interest for hosting topological surface states, the compounds $Bi_4Se_3$, i.e. $(Bi_2)(Bi_2Se_3)$ and BiTe i.e. $(Bi_2)(Bi_2Te_3)_2$ have been isolated, crystals have been grown, and topological surface states have been observed.[52, 53] For $Bi_4Se_3$ (Fig. 8a), partial S substitution was first found to stabilize the structure and allow for crystal growth, but recently the pure Se variant has been isolated and studied. [54]

The space group of $Bi_4Se_3$ is again $R\bar{3}m$ (#166). The structural parameterization is given by $a = 4.27$ Å (note the intermediate $a$ value, between that of elemental Bi (4.55 Å) and $Bi_2Se_3$ (4.14 Å), this is an indication of the interlayer bonding in the infinitely adaptive series) and $c = 40.0$.[55] The atoms are found in special positions as follows: Bi(1) *6c* (0,0,0.1445); Bi(2) *6c* (0,0,0.287); Se(1) *6c* (0,0,0.417) and Se(2) *3a* (0,0,0). The layer stacking sequence in $Bi_4Se_3$ is –{[Bi-Bi][Se-Bi-Se-Bi-Se]}$_0$–{[Bi-Bi][Se-Bi-Se-Bi-Se]}$_{1/3}$–{[Bi-Bi][Se-Bi-Se-Bi-Se]}$_{2/3}$–.

BiTe has a complex-looking structure (Fig. 8b).[56] It is the $n = 1$, $m = 2$ member of the infinitely adaptive series in the $(Bi_2)_n(Bi_2Te_3)_m$ system. This compound crystallizes in the trigonal space group $P\bar{3}m1$ (#164) with $a = 4.423$ Å $c = 24.002$ Å.[56] Apparently the rhombohedral stacking variant is not known, but instead the following stacking sequence is observed: –{[Bi-Bi] [Te-Bi-Te-Bi-Te]-[Te-Bi-Te-Bi-Te]}$_0$–{[Bi-Bi][Te-Bi-Te-Bi-Te]-[Te-Bi-Te-Bi-Te]}$_{1/2}$–. The atoms are found in special positions as follows: Bi(1) *2c* (0,0,0.1242); Bi(2) *2d* (1/3,2/3,0.2908); Bi(3) *2d* (2/3,1/3, 0.4575); Te(1) *2d* (1/3, 2/3, 0.0552); Te(2) *2d* (2/3, 1/3, 0.2149) and Te(3) *2c* (0,0, 0.3687).

Although other members of the $[(Bi,Sb)_2]_n[(Bi,Sb)_2(Se,Te)_3]_m$ infinitely adaptive series are known, and can be expected to display topological surface states with somewhat different properties than those currently known, they have not yet been grown as large crystals and tested by spectroscopy for the existence of such states.

The GeTe:Bi$_2$Te$_3$ series

This family[57, 58] has two members that display topological surface states.[59-61] From the chemical perspective, it is interesting that formal $Ge^{2+}Te^{2-}$ oxidation states (as are found in GeTe) can be inferred for these compounds (along with $Bi^{3+}$) from the fact that the two members known to be TIs, $GeBi_2Te_4$ and $GeBi_4Te_7$, are semiconductors.[60] The crystal structures of these two compounds (Figs. 9a and 9b) show that there is significant mixing of Ge and Bi on the metal sites; given the apparent differences in size and electronegativity for Ge and Bi, these compounds provide an interesting lesson in the flexibility of some materials systems to form unexpected phases. The compound $PbBi_2Se_4$, where Pb/Bi mixing is less surprising, is also a member of this structural family;[62] it is isostructural with $GeBi_2Te_4$.

The crystal structures of these compounds are significantly more complex than the others that are currently under study in the field; a "septuple layer" building block is found, that is layers, of the type: [Te-(Bi,Ge)-Te-(Bi,Ge)-Te-(Bi,Ge)-Te]. These are simply stacked in a rhombohedral cell in $GeBi_2Te_4$ (and $PbBi_2Se_4$),[63] in a septuple layer building block arrangement that is fully analogous to the quintuple layer building block arrangement in Tetradymites: –[Te-(Bi,Ge)-Te-(Bi,Ge)-Te-(Bi,Ge)-Te]$_0$–[Te-(Bi,Ge)-Te-(Bi,Ge)-Te-(Bi,Ge)-Te]$_{1/3}$–[Te-(Bi,Ge)-Te-(Bi,Ge)-Te-(Bi,Ge)-Te]$_{2/3}$–.

The crystal structure of $GeBi_2Te_4$ has been refined to a high degree of precision.[63] This has allowed the observation of a significant number of antisite defects in the compound. It crystallizes in space group $R\bar{3}m$ (#166); the cell parameters are $a$ = 4.322 Å (essentially equal to that found in $Bi_2Te_3$ where it is 4.39 Å), and $c$ = 41.270 Å. There are four independent atom sites as follows: [Bi,Ge](1), in the middle of the septuple layer, position *3a* (0,0,0), occupied in a 0.5:0.5 ratio of Bi to Ge. The second metal atom site, [Bi,Ge](2), describes the outer two metal layers in the septuple layer sandwich, position *6c* (0,0,0.4273); it is occupied in a 0.65:0.25:0.1 ratio of Bi:Ge:Te. Thus this site has a very large number of Te antisite defects. These antisite defects balance the Bi antisite defects that are found on the Te layers. The Te atoms in the external layers (Te(2)) are in *6c* (0,0,0.1344). These sites have 3% Bi antisite defects: they are occupied in the ratio 0.97:0.03 Te:Bi. The Te atoms in the internal Te layers (Te(1)) display a larger fraction of antisite defects. They are in position *6c* (0,0,0.2903) with an occupancy of 0.93:0.07 Te:Bi. Determining site mixing occupancies to this kind of precision is usual for crystal structure determinations. It shows that the structure, like that of $Bi_2Te_3$, is very flexible to the occurrence of both Bi and Te antisite defects. Thus, although the compound has been made either p-type or n-type through variations in composition, there is not a strong driving force for maintaining stoichiometry or site occupancy, and therefore it will likely be difficult to prepare with a very low carrier concentration. The crystal structure for isostructural $PbBi_2Se_4$, for which TI surface states have apparently also been observed has not been determined to a similar degree of precision and thus there is no crystallographic information available relevant to the defect chemistry.[64] In that case, the Pb and Bi are (reasonably) assumed to be randomly distributed on the metal sites. The crystal structure parameterization is: $a$ = 4.160 Å, $c$ = 39.200 Å, [Bi,Pb](1) *3a* (0,0,0); [Bi,Pb](2) *6c* (0,0,0.428); Se(1) *6c* (0,0,0.286); Se(2) *6c* (0,0,0.139).[62]

$GeBi_4Te_7$ and its Pb analog, $PbBi_4Te_7$, are semiconductors and host topological surface states,[60, 61] with a particularly interesting and complex crystal structure.[65] If the formula of a quintuple Tetradymite-type layer is generally written as $M_2Te_3$, and the formula of a septuple $GeBi_2Te_4$-type layer is similarly written as $M'_3Te_4$, then $GeBi_4Te_7$ is the n=1 m=1 member of a $(M_2Te_3)_n(M'_3Te_4)_m$ homologous series. The compound crystallizes in space group $P\bar{3}m1$ (#164), with cell parameters $a$ = 4.352 Å and $c$ = 23.925 Å. The metal sites are occupied in the statistical Ge:Bi ratio of 0.2:0.8. The two separate types of layers results in a large number of independent atomic sites. The positions are: [Bi,Ge](1), *1a* (0,0,0); [Bi,Ge](2) *2d* (1/3,2/3,0.838); [Bi,Ge](3) *2d* (1/3,2/3,0.584); Te(1) *2d* (1/3,2/3,0.074); Te(2) *2c* (0,0,0.232); Te(3) *2d* (1/3,2/3,0.342); and

Te(4) *1b* (0,0,0.5). Like in the $(Bi_2)_n(Bi_2Se_3)_m$ homologous series, it can be expected that cleavage of crystals of $GeBi_4Te_7$ will expose two different types of surfaces, a $Bi_2Te_3$-like surface and a $GeBi_2Te_4$-like surface. This has been established for the case of $PbBi_4Te_7$.[61] Larger *m* and *n* members of this family are known[55], but they have so far not been studied as candidates for TIs.

Topological Crystalline Insulators: SnTe, $Pb_{1-x}Sn_xTe$ and $Pb_{1-x}Sn_xSe$

The compound SnTe was the first predicted "topological crystalline insulator" (TCI), that is, a compound with metallic surface states whose stability is guaranteed ("protected" in the language of physics) by an element of crystal symmetry rather than by the usual case (so far) of what physicists refer to as time reversal symmetry.[15, 66] SnTe crystallizes in the space group $Fm\bar{3}m$ (#225), with the familiar rock salt structure composed of an FCC lattice of Te atoms, with Sn atoms filling all the octahedral voids.[67] Te and Sn occupy the Wyckoff sites *4a* and *4b*, respectively. The formal charges are $Sn^{2+}$ and $Te^{2-}$, implying from the high symmetry that the lone pair of s electrons on the Sn atom is not stereochemically active. In this particular system, the topological surface states are protected by the {110} family of mirror planes: the topological surface states are only observed on surfaces that are perpendicular to one of the {110} mirrors.[66]

SnTe has a direct band gap at the L point (the (0.5, 0.5, 0.5) point of the bulk Brillouin zone).[68] The electronic band structure at the L point is inverted (Sn states dominate the top of the valence band while Te states dominate the bottom of the conduction band) in SnTe and trivial (i.e., the normal case) in PbTe, which also has the rocksalt structure. This has been shown to result in a "topological phase transition" of the surface states from topological to trivial character in the $Pb_{1-x}Sn_xTe$ series,[66] which forms a solid solution for all *x*.[69] The lattice parameter is measurably different for the two end members. SnTe has a lattice parameter of 6.32 Å, with a Sn-Te bond length of 3.16 Å, while PbTe has a lattice parameter of 6.462 Å, with a Pb-Te bond length of 3.231 Å.[70, 71] The lattice parameter for $Pb_{1-x}Sn_xTe$ has a small positive deviation from Vergard's law; it is almost but not precisely linear in $x$.[72, 73] The electronic states are incredibly sensitive to lattice parameter – a 3% reduction in lattice parameter of PbTe would, it's predicted, drive it to a topological state.[66] The optimal material for study of the topological surface states in this system has been shown to be $Pb_{0.6}Sn_{0.4}Te$.[72, 73]

SnTe goes through a ferroelectric phase transition on cooling. The temperature of the transition is dependent on carrier concentration, and is accompanied by a cubic to rhombohedral structural distortion.[74] The phase transition maintains the basic crystal structure, but there is a distortion along the [111] direction, with the rhombohedral angles being slightly less than 90 degrees. This distortion is created by the Sn atoms moving slightly off center in the Te octahedra, which leads to the ferroelectric phase.[75] The temperature of the phase transition decreases with increasing carrier concentration; for SnTe with a p-type carrier concentration of $1.2 \times 10^{20}$ $cm^{-3}$, the phase transition occurs at 97.5 K; it is suppressed to 50 K at $5 \times 10^{20}$ $cm^{-3}$; and finally, the phase transition is not observed at a carrier concentration of $1.3 \times 10^{21}$ $cm^{-3}$.[75] Meanwhile, doping with Pb lowers the transition temperature; by $Pb_{0.6}Sn_{0.4}Te$ the phase transition is completely suppressed regardless of carrier concentration.[76] Thus the material best for the study of the topological surface states at low temperature, $Pb_{0.6}Sn_{0.4}Te$, is clearly in the stability range for the cubic phase at low temperatures.

The system $Pb_{1-x}Sn_xSe$ also has a topological phase transition, though neither end member is a TCI. PbSe crystallizes in the rock salt structure, with a lattice parameter of 6.128 Å. However, SnSe crystallizes in the orthorhombic space group Pnma (#62), and has a markedly different crystal structure; it is a layered compound with (SnSe) layers.[77] In these layers, Sn is coordinated to five Se atoms. This coordination is such that all five Sn-Se bonds point downwards, and the Sn atom is "exposed" from above, a consequence of the stereochemistry of the Sn lone pair. Within the layers, the configuration of the Sn-Se bonds alternates from up to down in one direction, and does not alternate in the other. Due to this marked structural difference, SnSe has a maximum solubility of 40% in PbSe; the $Pb_{1-x}Sn_xSe$ series maintains the rock salt structure only for $x < 0.4$. Given this, and that PbSe is topologically trivial, only compounds with x > 0 (the exact lower bound has not yet been determined) and x < 0.4 display the TCI surface states.[76, 78]

**Defect Chemistry**

Optimizing the bulk electronic properties of the topological insulators, which are small band gap semiconductors, relies on understanding their defect chemistry. The number of acceptor and donor defects should be balanced and minimized to make the bulk carrier concentration as low as possible, so that the interference of the bulk carriers with the surface carrier transport is small. Carrier concentrations in the $10^{16}$ - $10^{17}$ $cm^{-3}$ range or below are strongly preferred, though having orders of magnitude higher carrier concentrations does not hamper detection of the surface states by spectroscopic methods such as angle resolved photoelectron spectroscopy (ARPES). Low bulk carrier mobility is also favorable for surface state studies. However, both of these bulk characteristics should be achieved while inducing as few charged defects on the atomic layers supporting the topological surface states as possible. For two of the TI families, namely the Tetradymites and the rock salt structure TCIs the defect chemistry has been studied and understood by past researchers, in the context of conventional semiconductor studies. The defect chemistry for these two systems is described briefly in this section.

In the Tetradymite family, the strong prevalence of n-type behavior in undoped $Bi_2Se_3$ crystals arises from the presence of a high concentration of selenium vacancies, which are electron donors.[79] Using standard Kroeger-Vink notation, the formation of these defects can be described quantitatively by the reaction $Se_{Se} \rightarrow V_{Se}^{\cdot\cdot} + Se(g) + 2e'$, which denotes that a selenium atom sitting in the correct site in the crystal structure can escape as a vapor and thus leave behind a doubly-positively charged vacancy and two conduction electrons. The energy of formation of these vacancies is so small, approximately 0.5 eV, that their concentration is approximately $10^{20}$ cm$^{-3}$ at room temperature in $Bi_2Se_3$ prepared from stoichiometric melts by normal crystal growth processes such as the Bridgman method.[79, 80] Thus, there is a correspondingly large bulk concentration of n-type carriers in this material and the electrical conductivity of pure bulk bismuth selenide is far from insulating. The simplicity of the surface state in this material and the fact that its surface state Dirac point is at an energy that is distinctly different from that of any of the bulk carriers have made it the focus of much experimental study by physicists. However, after four years of intensive effort as of this writing, it has not yet been made as a bulk material or thin film with a sufficiently small number of defects to be insulating. It has been difficult to find suitable hole-dopants for compensating the Se vacancies to sufficiently lower the bulk carrier concentration. It is possible to make $Bi_2Se_3$ p-type by doping with very small amounts of very electropositive elements such as Ca substituted in place of Bi,[81] i.e., $Ca_{Bi2Se3} \rightarrow Ca_{Bi}' + h^{\cdot}$. Although this kind of substitution can be used to reduce the bulk carrier concentration to low levels, the strong effective charge of the electropositive Ca dopant is believed at this time to significantly limit the surface state mobility.

The defect chemistry of undoped $Bi_2Te_3$ also involves a large number of anion vacancies, but there is a higher energy of formation of the Te vacancy (calculated to be greater than 0.6 eV)[80] and therefore a smaller concentration of $V_{Te}^{\cdot\cdot}$ defects (Te$_{Te}$ → $V_{Te}^{\cdot\cdot}$ + Te($g$) + 2$e$') in this material than for $V_{Se}^{\cdot\cdot}$ in Bi$_2$Se$_3$. However, crystals of Bi$_2$Te$_3$ also contain a large concentration of singly-ionized Bi$_{Te}$' antisite defects. These occur to a larger extent in this compound than in Bi$_2$Se$_3$ because the more similar cation and anion electronegativities in Bi$_2$Te$_3$ means that they less strongly prefer one site over another in the structure. The equation for the creation of this defect is: Bi$_{Bi}$ → Bi$_{Te}$' + h$^{\cdot}$ + Te($g$) and its formation energy is similar to that of $V_{Te}^{\cdot\cdot}$ (approx. 0.5 eV).[82] These antisite defects complicate the picture by doping holes into the material and competing with vacancies for dominance. In fact, crystals grown from stoichiometric melts of Bi$_2$Te$_3$ are heavily doped p-type. The carrier type can be inverted to achieve heavily doped n-type crystals via Bi-poor/Te-rich growth conditions (the crossover occurs around 63 at. % Te).[83] However, the growth of highly compensated, low carrier concentration, high bulk resistivity crystals of Bi$_{2-x}$Te$_{2+x}$ has so far been elusive.

Antisite defects are even more pronounced in Sb$_2$Te$_3$ than they are in Bi$_2$Te$_3$ due to the stronger similarity between the electronegativities of Sb and Te. In fact, this material is so heavily doped p-type by native defects that n-type crystals have never been reported for any kind of dopant. The formation energy of Sb$_{Te}$' defects is only 0.35 eV, and their concentration in Sb$_2$Te$_3$ is almost 10 times that of Bi$_{Te}$' defects in Bi$_2$Te$_3$.[84] This compound can be alloyed with the above Tetradymites, and they with each other, to manipulate carrier concentrations, but these are considered as substitutional alloys rather than doped semiconductors. Sb$_2$Se$_3$ forms a structure different from that of the Tetradymites and so its defect chemistry will not be considered here.

Alloying Bi$_2$Te$_3$ and Bi$_2$Se$_3$ can form two phases of particular interest, namely Bi$_2$Te$_2$Se and Bi$_2$TeSe$_2$ as described above.[31] Both of these and their pseudo-ternary alloys including

Sb$_2$Te$_3$ have been widely studied as potential thermoelectric materials.[85] However, the former is of greater interest from the point of view of topological insulators because native crystals of Bi$_2$Te$_2$Se can be grown reliably with lower carrier concentrations and much higher electrical resistivities than the parent binary compounds.[31, 86] The primary defects in this material are charge-neutral Se-Te antisite defects, and the concentration of free carriers is influenced by Bi$_{Te}$' and $V_{Te}$·· defects, which dope holes and electrons, respectively. A careful balance of these competing defects can be achieved by growing crystals of "Bi$_2$Te$_2$Se" from a melt of slightly excess Bi (~2%) to produce samples with carrier concentrations that are significantly reduced from that of stoichiometric samples.

Both PbSe and PbTe exhibit similar behavior to Bi$_2$Te$_3$ in that crystals of these compounds contain two competing defects. Thus, they can display either n- or p-type semiconducting behavior, depending on whether an excess of cations or anions is present, respectively.[87-89] However, the defect chemistry differs slightly between PbSe and PbTe. On the one hand, an excess of anions produces the same result in both PbSe and PbTe; namely, it leads to a high concentration of Pb vacancies and thus creates holes: Pb$_{Pb}$ → $V_{Pb}$'' + 2h· + Pb(g). On the other hand, the free electrons in crystals of Pb-rich PbSe are generated by Pb interstitial defects: Pb(g) → Pb$_i$·· + 2e', while PbTe crystals with excess Pb instead contain Te vacancies: Te$_{Te}$ → $V_{Te}$·· + 2h' + Te(g). In all these cases, each point defect is doubly ionized and generates two free carriers of the appropriate type. In PbSe, typical values for $n$ are between $10^{18}$ and $10^{19}$ cm$^{-3}$ and $p$ between 3 and 6.5x$10^{18}$ cm$^{-3}$.[90] In PbTe, $n$ and $p$ are both usually found in the range of $10^{17}$ to $10^{18}$ cm$^{-3}$. PbTe is not easily made n-type.[80, 91, 92]

The predominant defect in SnTe is the Sn vacancy, regardless of Sn or Te supersaturation. This defect also generates two holes (Sn$_{Sn}$ → $V_{Sn}$'' + 2h· + Sn(g)), and leads to a concentration of free hole carriers in the bulk on the order of $10^{20}$ to $10^{21}$ cm$^{-3}$.[93] Similar to the case of Sb$_2$Te$_3$, n-type crystals of SnTe have not been reported. PbTe-SnTe alloys can be prepared across the entire solid solution range, as described above, and generally show p-type carrier concentrations between those of the end members.[92]

**Conclusions – Looking for Topological Insulators**

To truly predict whether a particular compound will display topological surface states requires the calculation of good quality bulk and surface band structures, and a theoretical analysis of the parity of the electronic bands. Materials chemists may thus have to rely on the published predictions of theoretical or computational physicists if they wish to find new TIs, but that comes with its own set of problems (see below). A simple set of chemical characteristics providing general guidelines can be deduced, however, from the known topological insulators. Firstly, heavy elements must be involved, to give significant amounts of spin orbit coupling. This is satisfied by elements from the 5s5p – 6s6p family, although the compound need not consist exclusively of such elements (e.g., GeBi$_2$Te$_4$). Secondly, in order for the most basic bulk electronic characteristics to be present, the mixing of electronic states dominated by the more electropositive element into the valence band and electronic states dominated by the more electronegative element into the conduction band must be present. This implies that the differences in electronegativities among the elements whose states are near the Fermi energy must be small. The presence of the spin orbit coupling impacts the energies of the states at the bottom of the conduction band and the top of the valence band, deforming them from the usual parabolic energy vs. electron wave vector relationship found in most bulk materials. This spin orbit coupling can result in the opening of a band gap in cases where the material would have been a semimetal. The net result is that the highest chance of finding topological surface states can be found for bulk materials that are either semimetals or semiconductors with band gaps less than about 0.4 eV. Finally there is the issue of crystal symmetry. For classical topological insulators to occur, the surface state bands must cross from the bulk valence band to the bulk conduction band an odd number of times over the whole Brillouin zone. A single crossing of the gap is possible for any material with axial symmetry if the conduction band minimum and valence band maximum are at the center of the Brillouin zone. This odd-crossing criterion can be satisfied for electrons of general wavevector if the symmetry of the surface of interest has three-fold symmetry, such as is found for bulk materials in Laue classes P3 or R3, e.g. as is found for the Tetradymites.

With these general characteristics in mind, solid state chemists can reasonably guess which compounds and which of their surfaces may host topological surface states even without the benefit of detailed predictions. The fact that electronic structure calculations are sufficiently user-friendly that chemists can run them to test their ideas is very helpful. Only bulk calculations looking for the above characteristics are needed to support a basic guess, though they will not be sufficient to prove the topological character of the surface states – more sophisticated surface state calculations or experimental determination of the spin polarization of the surface states are needed.

Solid state chemists and materials scientists who are interested in growing crystals or fabricating thin films of theoretically predicted but currently unproven TIs can find many opportunities for discovery in published theoretical predictions. Caution is suggested, however, in that chemical common sense should be used to screen the predictions before making the significant time investment needed in growing films or crystals and testing them; factors such as the radioactivity and toxicity of the proposed compounds, predictions made for compounds that do not actually exist, predictions made for compounds where the symmetry or crystal structure employed in the calculation is different from what is present for the real material, predicted spin states that are impossible, predictions that obviously do not match previously measured properties, and incorrectly predicted magnitudes of band gaps have not prevented the publication of such predictions in high impact journals. Solutions to the calculation-related parts of the prediction issue have been proposed and implemented, and are worth understanding.[94] Chemists working in this area should not be surprised if on the one hand most physicists' eyes will glaze over when insufficient background is provided for them to appreciate the chemical or structural complexities present in your compound, but on the other hand will often find chemical ideas to be very interesting when properly explained. Finding common language is important for progress in this field, as it is in others; topological insulators are an excellent area for collaborative research between physicists and chemists, and will be of interest for years to come.

**Acknowledgements**

The authors' work on Topological Insulators has been supported by the US National Science Foundation MRSEC program grant DMR-0819860, by DARPA SPAWAR grant NN66001-11-1-4110. Our work on low temperature thermoelectrics, relevant to the discussion of the defect chemistry, has been supported by AFOSR grant FA9550-10-1-0533. The authors thank Jason Krizan of Princeton University for the photographs of the crystals in figures 2b and 3b, and Bin Liu of USTC for his help with the text. Structure figures were drawn employing the program VESTA.

**Figures**

**Fig. 1** (color on line) (a) The crystal structure of HgTe. The unit cell is outlined in solid lines. The Hg atoms and Te atoms are plotted as grey and blue spheres, respectively. Each Hg atom is coordinated by four Te atoms and a 3D corner sharing tetrahedra network is formed.

**Fig. 2** (color on line) (a) The crystal structure of elemental Bi and Sb. The unit cell is outlined in solid lines. Each unit cell is comprised of three Bi/Sb bi-layers with van der Waals interaction in between. Each Bi/Sb layer follows the ABCABC stacking format. (b) Basal plane cleavage surfaces of a $Bi_{0.9}Sb_{0.1}$ alloy crystal.

**Fig. 3** (color on line) (a) The crystal structure of $Bi_2Se_3$. The Bi atoms and Se atoms are plotted as purple and green spheres, respectively. The octahedra are in the same color scheme with the central atoms. The unit cell is outlined in solid lines. Each unit cell is comprised of three –[Se(2)-Bi-Se(1)-Bi-Se(2)]– quintuple layers. Each Bi atom is coordinated by six Se atoms and the $BiSe_6$ octahedra share edges within a quintuple layer. (b) A section of a crystal of $Bi_2Se_3$ grown by the Bridgeman method.

**Fig. 4** (color on line) The progression of structures in the Tetradymite series, shown for one quintuple layer sandwich: $Bi_2Te_3$-$Bi_2Te_2Se$-$Bi_2TeSe_2$-$Bi_2Se_3$. The Bi, Se, and Te atoms are plotted as purple, green, and blue spheres, respectively. Se atoms strongly prefer the central layer and therefore $Bi_2Te_2Se$ forms a crystallographically ordered structure, in which Te atoms and Se atoms are well separated.

**Fig. 5** (color on line) The crystal structure of the $Bi_2(Te_{0.8}S_{0.2})_2S$ Tetradymite. The unit cell is outlined in solid lines. The Bi, Te, and S atoms are plotted as purple, blue, and yellow spheres, respectively. The central layer within a quintuple layer is solely occupied by S atoms, while the outer layers are mixed with 80%Te/20%S. Compared with Te atoms, S atoms are closer to the Bi layer as a result of their smaller atomic radii.

**Fig. 6** (color on line) The crystal structure of TlBiSe$_2$. The unit cell is outlined in solid lines. The Tl, Bi, and Se atoms are plotted as bronze, purple, and green spheres, respectively. Each Tl/Bi atom is coordinated by six Se atoms, and a 3D edge sharing octahedra network is formed. Tl and Bi layers alternate ideally in this depiction, while their real ordering in crystallography is still unclear.

**Fig. 7** (color on line) The infinitely adaptive series (Bi$_2$)$_n$(Bi$_2$Te$_3$)$_m$. The Bi atoms and Te atoms are plotted as purple and blue spheres, respectively. The (Bi$_2$) and (Bi$_2$Te$_3$) building blocks are depicted as purple and blue rectangles. The unit cell is outlined in solid lines. Different stacking sequences result in varied stoichiometries in the infinitely adaptive series.

**Fig. 8** (color on line) The crystal structures of (a) Bi$_4$Se$_3$ and (b) BiTe (same color scheme). The unit cells are outlined in solid lines. The Bi$_2$ bi-layer and Bi$_2$X$_3$ quintuple layer building blocks are depicted in ball-and-stick and polyhedral forms, respectively. The crystal structure of BiTe is shown with c-axis doubled to manifest the stacking sequence.

**Fig. 9** (color on line) The crystal structures of (a) GeBi$_2$Te$_4$ and (b) GeBi$_4$Te$_7$. The Bi/Ge atoms are plotted as pink spheres. The unit cells are outlined in solid lines. A GeBi$_2$Te$_4$ unit cell is composed of three (Bi,Ge)$_3$Te$_4$ septuple layers, while a GeBi$_4$Te$_7$ unit cell is composed of a (Bi,Ge)$_3$Te$_4$ septuple layer and a (Bi,Ge)$_2$Te$_3$ quintuple layer.

92. C. R. Hewes, M. S. Adler and S. D. Senturia, *J. Appl. Phys.*, 1973, **44**, 1327.

93. R. F. Brebrick, *J. Phys. Chem. Solids*, 1963, **24**, 27.

94. J. Vidal, X. Zhang, L. Yu, J. W. Luo and A. Zunger, *Phys. Rev. B*, 2011, **84**, 041109.

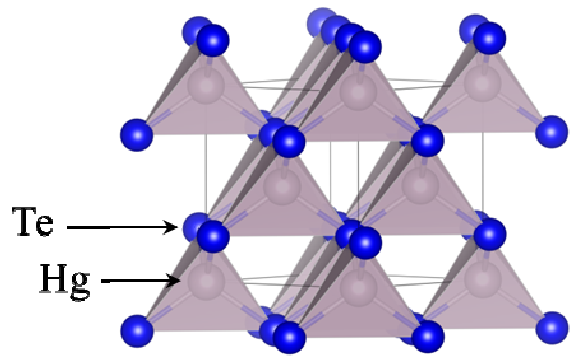

**HgTe**

**Fig. 1**

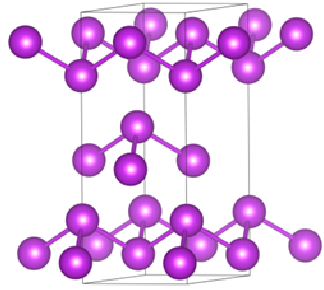

**Fig. 2a**

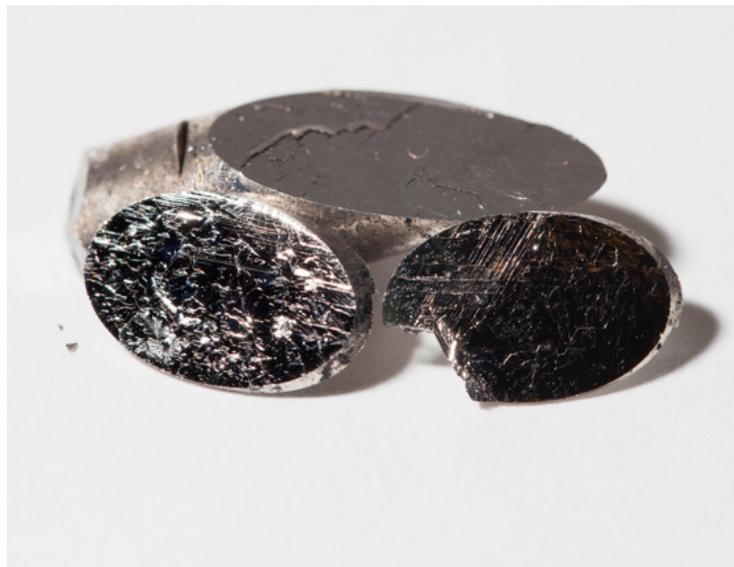

**Fig. 2b**

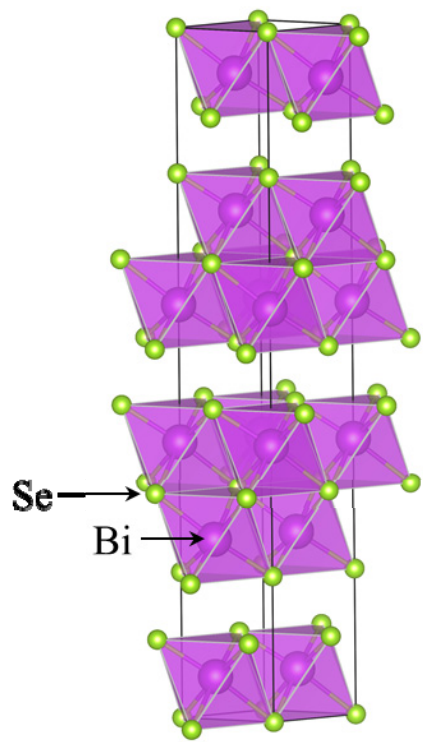

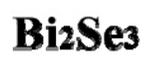

Fig. 3a

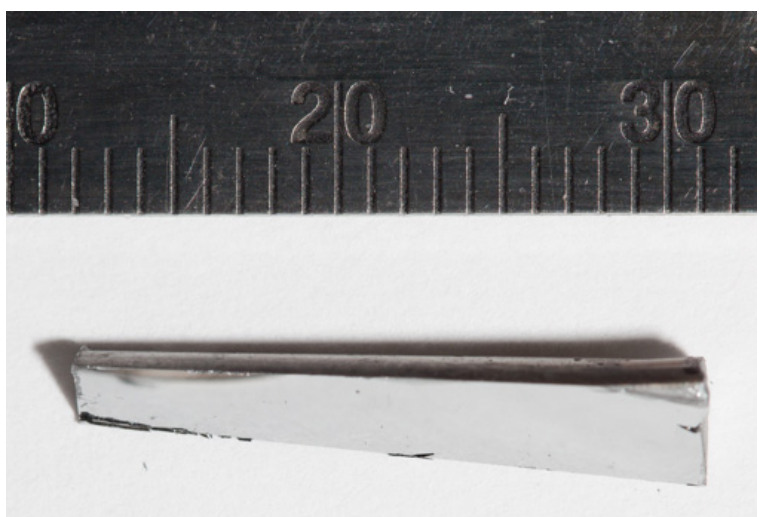

Fig. 3b

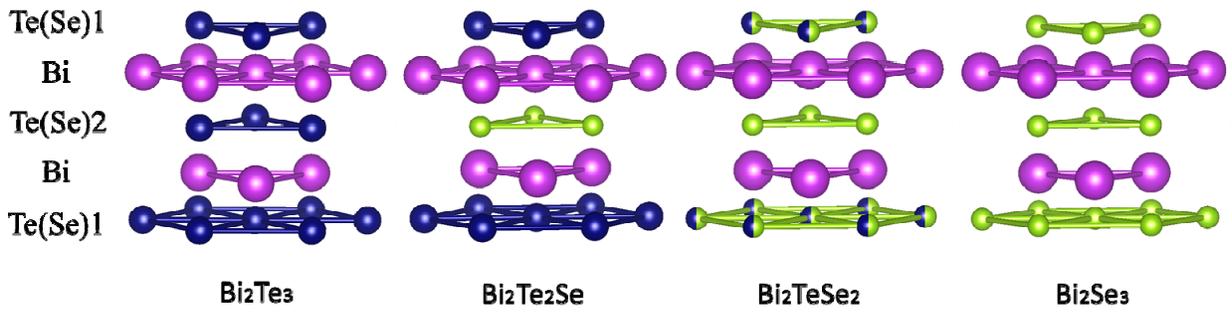

**Fig. 4**

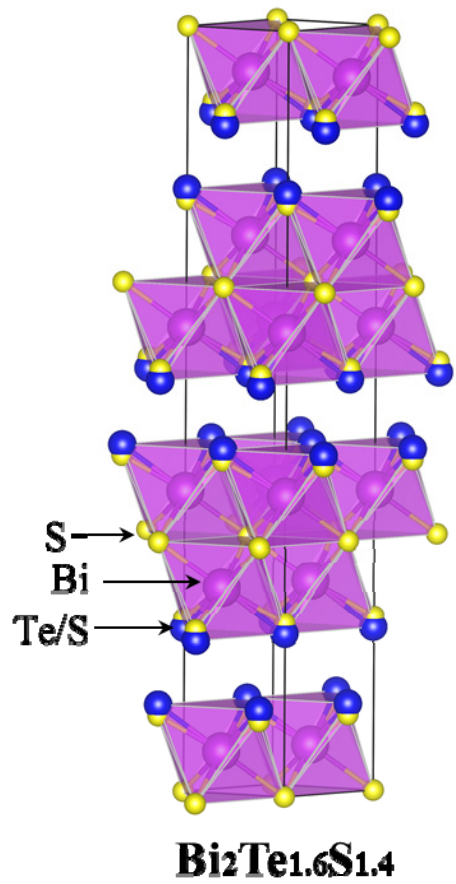

**Fig. 5**

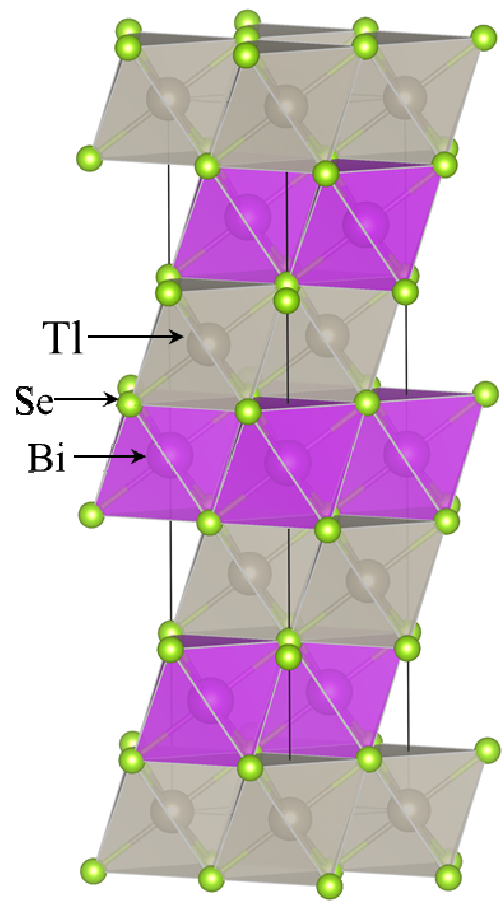

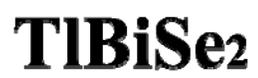

**Fig. 6**

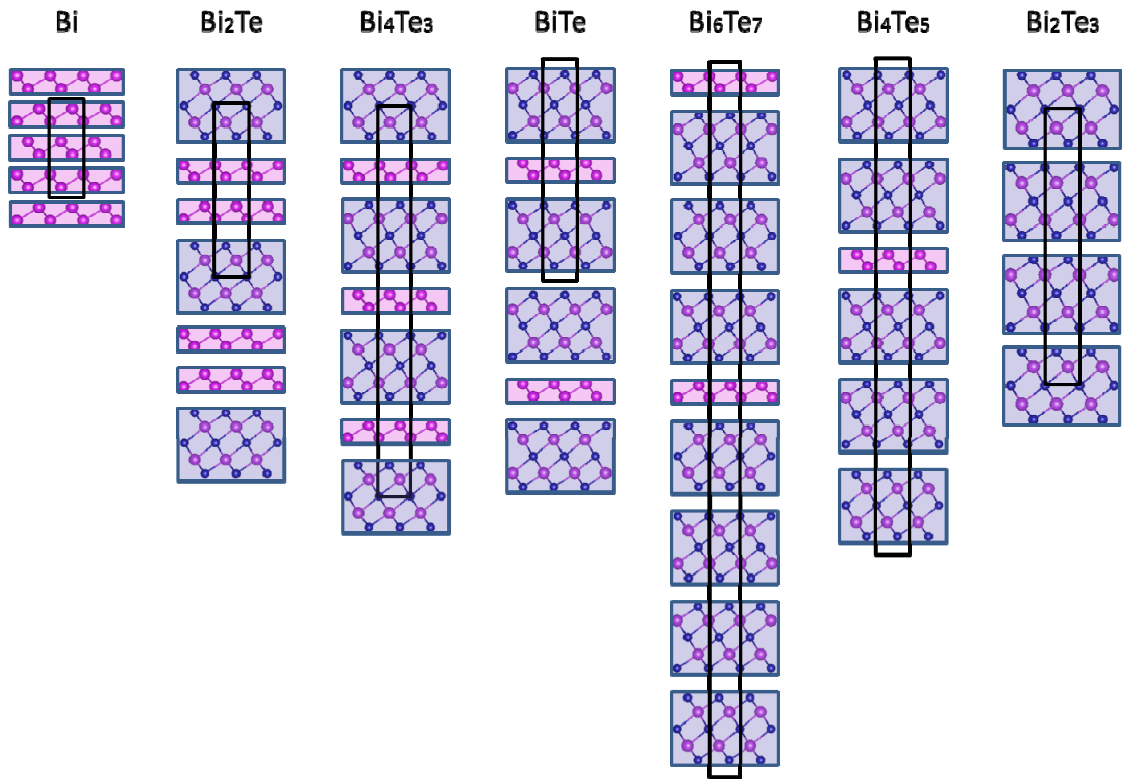

**Fig. 7**

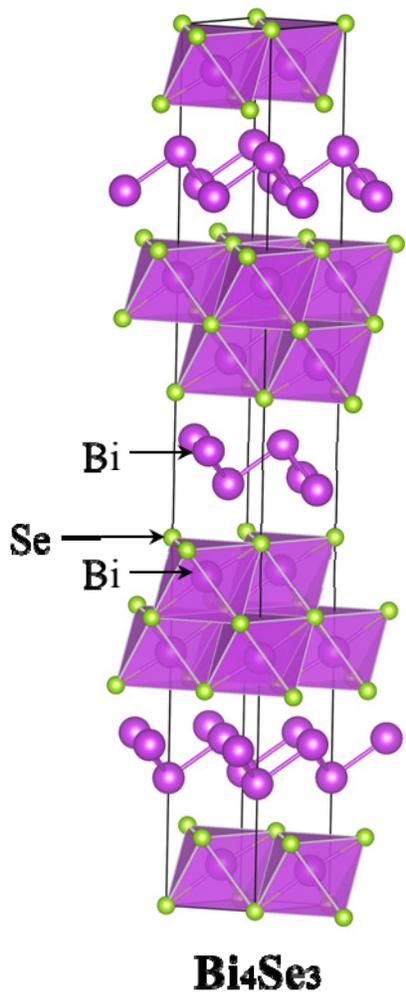

**Fig 8a**

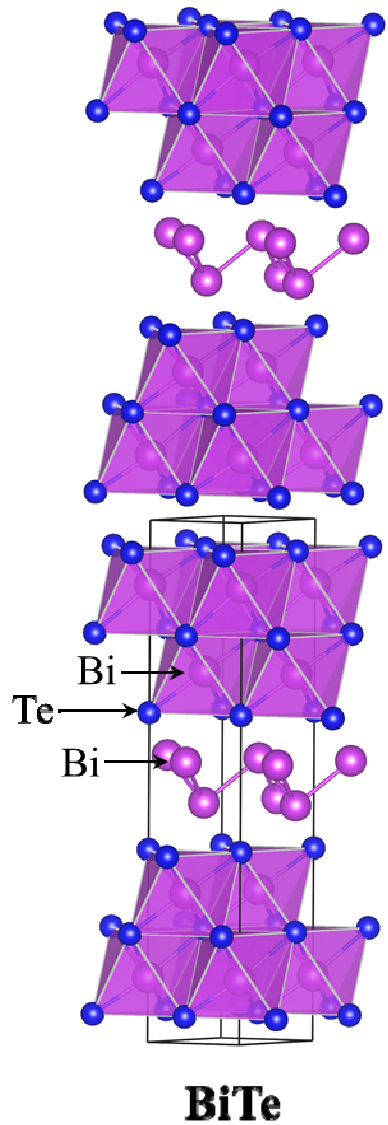

**Fig. 8b**

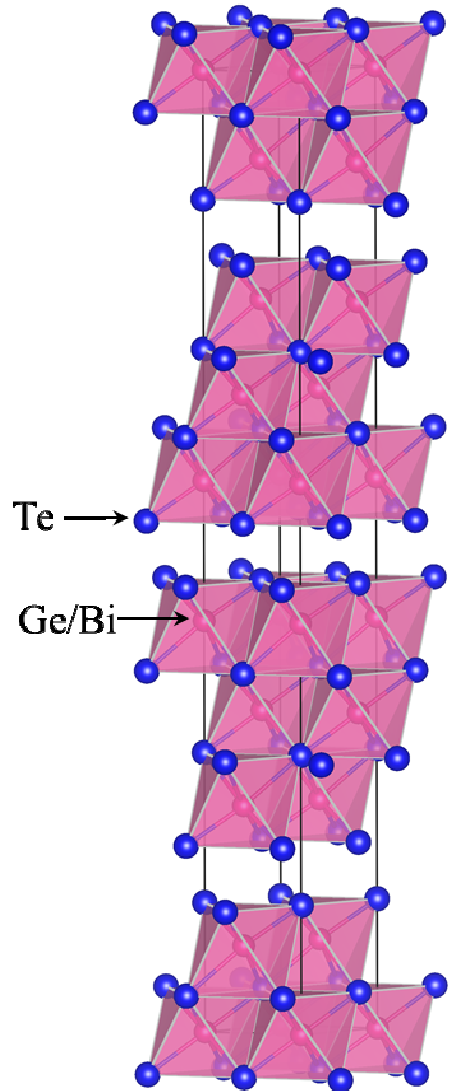

**Fig. 9a**

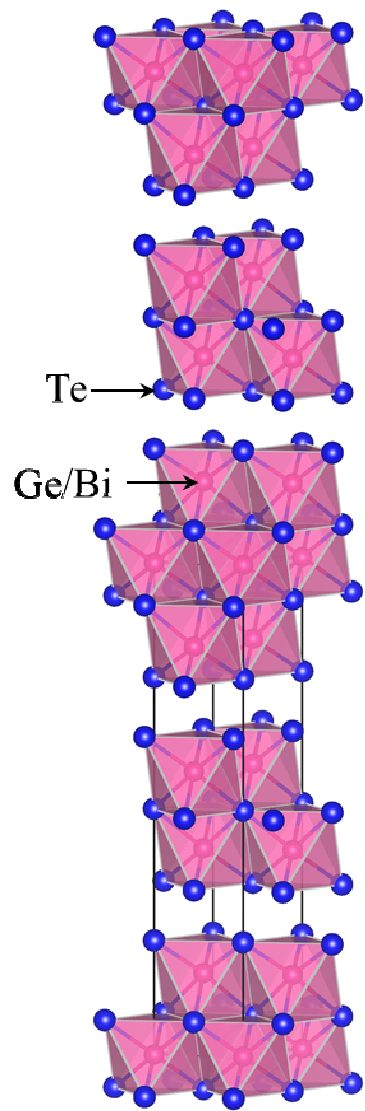

**GeBi$_4$Te$_7$**

**Fig. 9b**